\documentclass{article}

\usepackage{graphicx}
\usepackage{chapterbib}
\usepackage{bm}%
\usepackage{braket,amsmath,mathabx}
\usepackage{newfloat}
\usepackage{multicol}
\usepackage{setspace}

\DeclareFloatingEnvironment[
    fileext=los,
    name=Figure,
    placement=tbhp,
    within=none,
]{suppfig}

\usepackage{times}
\usepackage[pdftex,bookmarks=true,bookmarksopen,bookmarksnumbered,
                colorlinks,
                linkcolor=blue,
                citecolor=blue]{hyperref}
\usepackage{caption}

\newenvironment{sciabstract}{%
\begin{quote} \bf 
}
{\end{quote}}

\newcounter{lastnote}

\newenvironment{addendum}{%
    \setlength{\parindent}{0in}%
    \small%
    \begin{list}{Acknowledgements}{%
        \setlength{\leftmargin}{0in}%
        \setlength{\listparindent}{0in}%
        \setlength{\labelsep}{0em}%
        \setlength{\labelwidth}{0in}%
        \setlength{\itemsep}{12pt}%
        }
    }
    {\end{list}\normalsize}

\newcommand*{\addendumlabel}[1]{\textbf{#1}\hspace{1em}}

\title{A Dressed Spin Qubit in Silicon}

\author
{Arne Laucht,$^{1,\ast}$ Rachpon Kalra,$^1$ Stephanie Simmons,$^1$ Juan P. Dehollain,$^1$ \\Juha T. Muhonen,$^1$ Fahd A. Mohiyaddin,$^1$ Solomon Freer,$^1$ \\Fay E. Hudson,$^1$ Kohei M. Itoh,$^2$ David N. Jamieson,$^3$ Jeffrey C. McCallum,$^3$ \\Andrew S. Dzurak,$^1$ and A. Morello$^{1,\ast}$\\
\\
\normalsize{$^{1}$Centre for Quantum Computation and Communication Technology, }\\
\normalsize{School of Electrical Engineering and Telecommunications, }\\
\normalsize{UNSW Australia, Sydney, New South Wales 2052, Australia}\\
\normalsize{$^{2}$School of Fundamental Science and Technology, Keio University, }\\
\normalsize{3-14-1 Hiyoshi, 223-8522, Japan}\\
\normalsize{$^{3}$Centre for Quantum Computation and Communication Technology, }\\
\normalsize{School of Physics, University of Melbourne, Melbourne, Victoria 3010, Australia}\\
\\
\normalsize{$^\ast$Corresponding authors. E-mail: a.laucht@unsw.edu.au and a.morello@unsw.edu.au.}
}

\date{}

\begin{document}
\spacing{1.15}
\maketitle
\baselineskip16pt

\begin{sciabstract}
Coherent dressing of a quantum two-level system provides access to a new quantum system with improved properties - a different and easily tuneable level splitting, faster control, and longer coherence times. In our work we investigate the properties of the dressed, donor-bound electron spin in silicon, and probe its potential for the use as quantum bit in scalable architectures. The two dressed spin-polariton levels constitute a quantum bit that can be coherently driven with an oscillating magnetic field, an oscillating electric field, by frequency modulating the driving field, or by a simple detuning pulse. We measure coherence times of $T_{2\rho}^*=2.4$~ms and $T_{2\rho}^{\rm Hahn}=9$~ms, one order of magnitude longer than those of the undressed qubit. Furthermore, the use of the dressed states enables coherent coupling of the solid-state spins to electric fields and mechanical oscillations.
\end{sciabstract}

\begin{figure*}[!t]
\begin{center}
\includegraphics[width=1\textwidth]{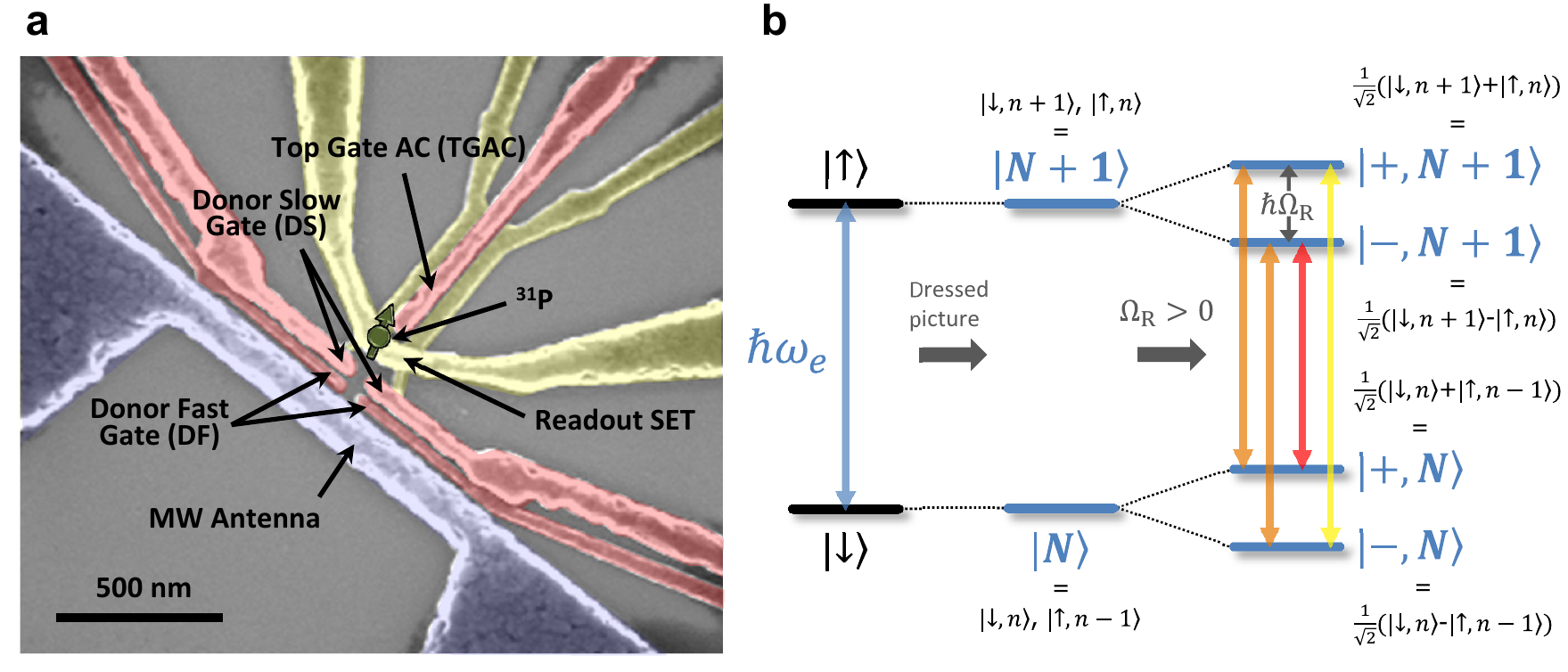}
\caption{\label{figure01} \textbf{Sample design and energy level diagram.} 
\textbf{a}, Scanning electron microscope image of a device, similar to the one used for measurements. Highlighted are the positions of the $^{31}$P donor, the donor gates, the microwave (MW) antenna, and the single electron transistor (SET) for spin readout.
\textbf{b}, Energy level diagram of the electron spin subsystem in the spin picture and the dressed picture.
}
\end{center}
\end{figure*}

\newpage
\begin{multicols}{2}
\setstretch{1.15}
Coherent dressing of a quantum two-level system has been demonstrated on a variety of systems, including atoms~\cite{Mollow1969}, self-assembled quantum dots~\cite{Xu2007}, superconducting quantum bits~\cite{Baur2009}, and NV centres in diamond~\cite{London2013}. In this context ``dressing'' means that an electromagnetic driving field coherently interacts with the quantum system, so that the eigenstates of the driven system are the entangled states of the photons and the quantum system. For the case of a single spin in a static magnetic field $B_0$ that is driven with an oscillating magnetic field $B_1$, this means that the eigenstates are no longer the spin-up and spin-down states, but the symmetric and antisymmetric superpositions of these states with the driving field (comp. Fig.~\ref{figure01}\textbf{b}). These new eigenstates are often termed “polaritons” and for a single electron spin, the name “electron spin polariton” would be appropriate.

The motivation to take a two-level system into the dressed basis is multifaceted. The dressed system possesses a level splitting proportional to the driving strength, and can therefore be dynamically modified and tuned to be in resonance with other quantum systems~\cite{London2013,Hartmann1962,Cai2013}. Furthermore, the continuous driving decouples the spin from background magnetic field noise. Even in highly coherent systems such as trapped atoms, coherence times can be improved by two orders of magnitude by dressing the states~\cite{Timoney2011}. Finally, the change in quantization axis unlocks new ways of operating on a dressed qubit. A dressed qubit can be controlled by changes in the bare spin Larmor frequency, allowing manipulation using electric fields and strain via the hyperfine coupling in donor systems~\cite{Laucht2015,Dreher2011}. This opens up the possibility of coupling spins to the motion of a mechanical oscillator when the transition frequency between dressed states becomes comparable to the oscillator's frequency~\cite{Rabl2009}.

Because of these benefits, it has been proposed to use dressed states for quantum gates and memories for quantum computing~\cite{Timoney2011,Mikelsons2015,Cai2015}. In these proposals, the focus was on optically-dressed qubits in trapped atomic ions, because dressing a qubit fundamentally requires a high ratio between the driving field and the intrinsic resonance linewidth, most easily achieved in optics and trapped ions. Here, we demonstrate a microwave-dressed spin qubit in the solid state, using a highly coherent $^{31}$P donor bound electron in silicon~\cite{Pla2012}. We demonstrate that the use of dressed states, compared to bare spin states, has several advantages for quantum computation.

The $^{31}$P donor in silicon constitutes a natural two-qubit system, where both the electron (indicated with $\ket{\downarrow}$ or $\ket{\uparrow}$) and the nuclear ($\ket{\Downarrow}$ or $\ket{\Uparrow}$) spin states can be coherently controlled by a magnetic field $B_1$ oscillating at specific electron spin resonance (ESR) and nuclear magnetic resonance (NMR) frequencies. We fabricated a device that comprises a single $^{31}$P donor in an isotopically purified $^{28}$Si epilayer~\cite{Itoh2014}, implanted \cite{Jamieson2005} next to the island of a single-electron-transistor (SET). The SET is formed under an 8~nm thick SiO$_2$ layer by biasing a set of electrostatic gates (yellow in Fig.~\ref{figure01}\textbf{a}). The distance between donor and SET island is $\sim 20(5)$~nm~\cite{Laucht2015}, with a tunnel coupling of order 10~kHz. The device is cooled by a dilution refrigerator (electron temperature $T_{\rm el} \approx 100$~mK), and subject to a static magnetic field $B_0 = 1.55$~T applied along the [110] Si crystal axis. Due to the Zeeman effect the electrochemical potential $\mu$ of the donor electron depends on its spin state, with $\mu_{\uparrow} > \mu_{\downarrow}$. Another set of gates (pink in Fig.~\ref{figure01}\textbf{a}) is used to tune the electrochemical potentials of donor and SET island ($\mu_{\rm SET}$) to the readout position, where $\mu_{\uparrow} > \mu_{\rm SET} > \mu_{\downarrow}$ such that only an electron in the $\ket{\uparrow}$ state can tunnel out of the donor. The positive donor charge left behind shifts the electrochemical potential of the SET island and causes a current to flow, until a $\ket{\downarrow}$ electron tunnels back onto the donor. This spin-dependent tunneling mechanism is used to achieve high-fidelity, single-shot electron spin readout \cite{Morello2009,Morello2010}, as well as initialization of the donor electron spin into the $\ket{\downarrow}$ state. For coherent spin control, an oscillating magnetic field $B_1$ is delivered to the donor by an on-chip, broadband transmission line terminating in a short-circuited nanoscale antenna~\cite{Dehollain2013} (blue in Fig.~\ref{figure01}\textbf{a}). While manipulating the electron spin state, the gates are tuned such that $\mu_{\uparrow,\downarrow} < \mu_{\rm SET}$, to ensure that the electron cannot escape the donor.

\section{Dressing the electron spin - Rabi oscillations and Mollow triplets} 

Fig.~\ref{figure01}\textbf{b} shows the energy level diagram of the electron spin subsystem. Naturally, we would draw the energy levels in the $|{\uparrow\rangle}$\textemdash$|{\downarrow\rangle}$ basis which corresponds to the spin picture, with electron spin transition frequency $\hbar\omega_e$. Alternatively, we can draw them in the dressed $|{N+1\rangle}$\textemdash$|{N\rangle}$ basis, where $N$ is the total number of excitations in the system. The state $|{N\rangle}$ then consists of the two degenerate states $|{\downarrow,n\rangle}$ and $|{\uparrow,n-1\rangle}$, where $n$ is the number of resonant photons in the driving field. For a non-zero spin-photon coupling, the dressed levels are split into the entangled $|{+,N\rangle}=\tfrac{1}{\sqrt{2}}(|{\downarrow,n\rangle}+|{\uparrow,n-1\rangle})$ and $|{-,N\rangle}=\tfrac{1}{\sqrt{2}}(|{\downarrow,n\rangle}-|{\uparrow,n-1\rangle})$ states which differ in energy by the Rabi frequeny $\hbar\Omega_R=\hbar \frac{1}{2} \gamma_e B_1$, where $\gamma_e$ is the gyromagnetic ratio of the electron and $B_1$ is the amplitude of the oscillating driving field.

We can define the dressed spin states as the basis states of a new qubit, the dressed qubit. Here, the states $\ket{-}$ and $\ket{+}$ act as the computational basis states. Note that we have omitted $N$, the number of excitations in the system, as the electron is dressed by a classical driving field, where the number of photons $n$ is very large and its exact value is unimportant. We do, however, need to take into account the magnitude of $B_1$, which defines the electron Rabi frequency $\Omega_R$ and therefore the splitting between the $\ket{-}$ and $\ket{+}$ states.

\begin{figure*}[!t]
\begin{center}
\includegraphics[width=1\textwidth]{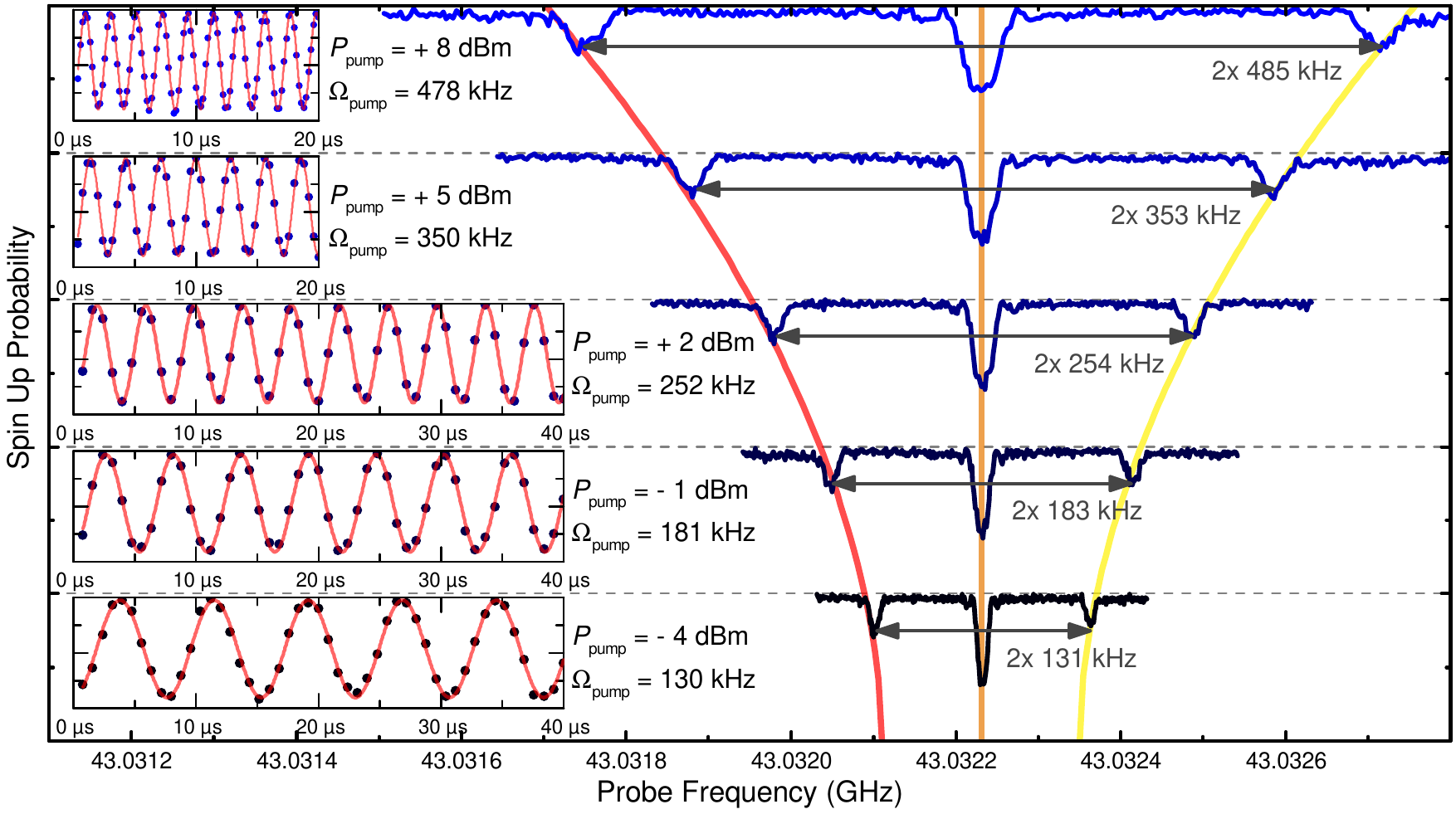}
\caption{\label{figure02} \textbf{Dressing the electron spin.} Mollow spectra of the dressed electron spin for different microwave powers. Here, a strong, resonant driving field $P_{\rm pump}$ is used to dress the spin state, while a weaker probe field $P_{\rm probe} = P_{\rm pump}-26$~dB is scanned in frequency to record the Mollow triplet spectra. (The curves are shifted in frequency to compensate the slight drift of the background magnetic field $B_0$.) The insets on the left show standard Rabi experiments for the same microwave powers (the vertical scale is the electron spin-up probability, from 0 to 1), with Rabi frequencies that match the splitting between the peaks of the Mollow triplet.
}
\end{center}
\end{figure*}

There are two ways to demonstrate creation of dressed states. The first is by performing coherent Rabi oscillations, which has been demonstrated on electron and nuclear spins in many material systems for ensembles~\cite{Abragam1961,Vandersypen2005} and single spins~\cite{Jelezko2004,Koppens2006,Press2008,Muhonen2014,Veldhorst2014}. The second is by measuring the Mollow triplet~\cite{Mollow1969}. This, however, is more difficult, as the Mollow spectrum is usually measured in transmission/absorption~\cite{Xu2007,Kroner2008,Baur2009} or fluorescence/scattering~\cite{Mollow1969,Wu1975,Astafiev2010} and requires a very sensitive optical detector and outstanding background suppression. Here, we implement a novel way to measure the Mollow spectrum, based on our high-fidelity control and spin readout. After initializing the donor electron in the $\ket{\downarrow}$ state, we use a resonant, high-power microwave (MW) pulse (pump pulse) to perform a controlled rotation of angle $(2\mathrm{n}+1)\cdot\pi$, which rotates the spin to the $\ket{\uparrow}$ state. The pump pulse serves to create the dressed states with splitting equal to the Rabi frequency $\Omega_{\rm pump}$. At the same time, a low-power microwave pulse (probe pulse) of the same length is scanned over the resonance to probe the response of the driven spin to different frequencies. 

We plot the results of this measurement in Fig.~\ref{figure02}. The different curves were recorded for different microwave powers. The pump power was increased from $P_{\rm pump}=-4$~dBm (at the source) to $P_{\rm pump}=+8$~dBm, while the probe power was always kept $26$~dB lower, i.e. $P_{\rm probe}=P_{\rm pump} - 26$~dB. The length of the two simultaneous pulses was chosen such that the resonant pump pulse results in a $21\pi$ rotation of the electron spin, whereas the probe pulse by itself would lead to $\approx\pi$ rotation when in resonance. When the probe frequency is scanned over the resonance, the effect it has on the driven spin nicely reproduces the expected Mollow triplet. The spectrum consists of a larger peak in the middle and two smaller peaks split off by the Rabi frequency $\Omega_{\rm pump}$, corresponding to the four different transitions indicated in red, orange and yellow in Fig.~\ref{figure01}\textbf{b}. When increasing the microwave power, this splitting increases due to the stronger drive, as expected. At the same time, the width of the peaks increases due to the higher probe power. The insets of Fig.~\ref{figure02} show standard Rabi experiments for the same microwave powers. Here, the dots correspond to the experimental result, while the red lines are fits to a sinusoid. The fits allow us to extract the Rabi frequency from the oscillations, in excellent agreement with the peak splitting of the Mollow triplet. In the supplementary material (see section \ref{SuppMollow}) we show additional data where the pulse length was reduced (see Fig.~\ref{figureS_pulselength}) and also where the ratio $P_{\rm probe}/P_{\rm pump}$ was changed (see Fig.~\ref{figureS_probepower}).

\section{\label{Init}Dressed qubit initialization and readout}

\begin{figure*}[!t]
\begin{center}
\includegraphics[width=1\textwidth]{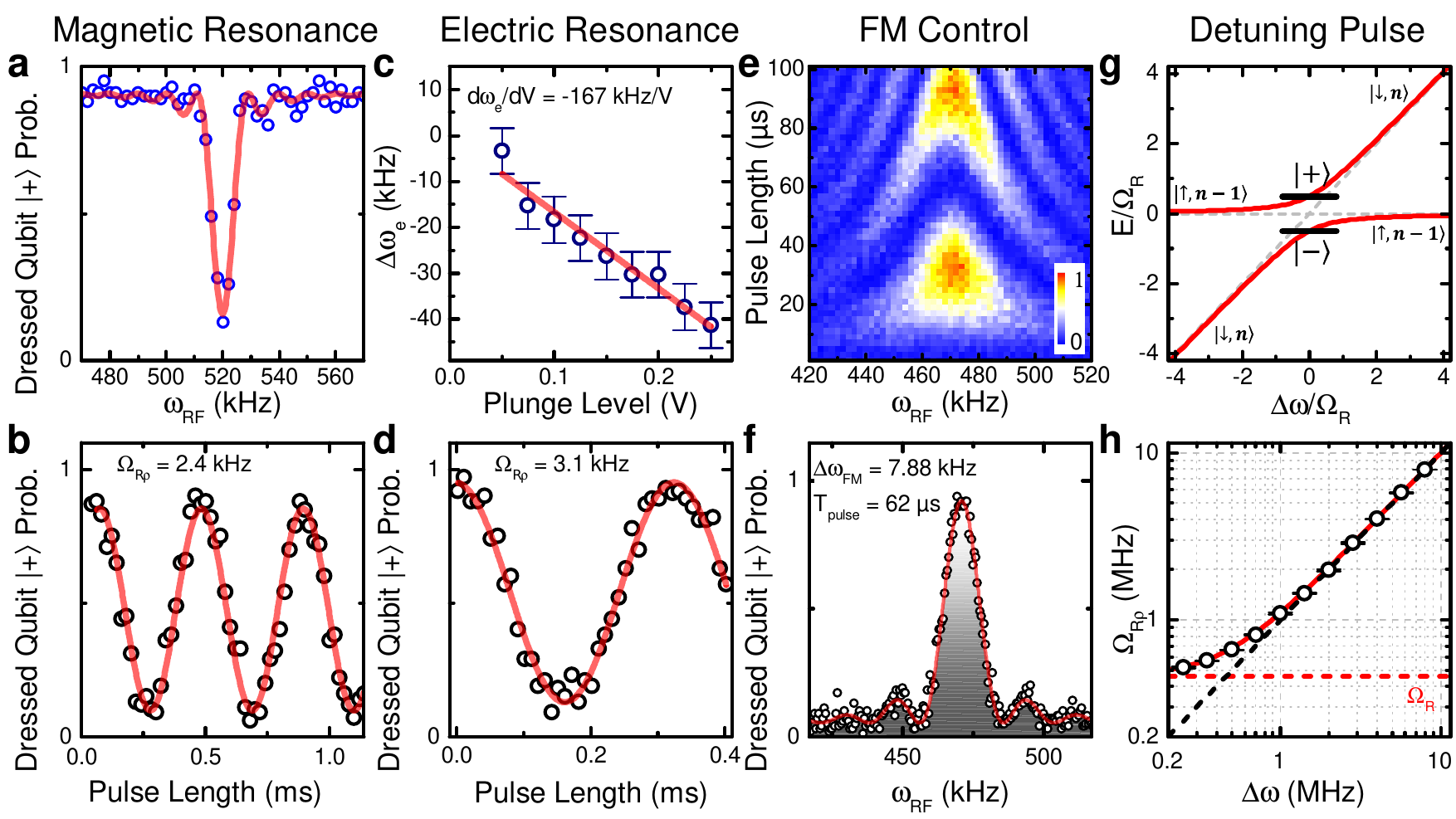}
\caption{\label{figure04} \textbf{Dressed qubit control.}
\textbf{a}, Dressed qubit spectrum and \textbf{b}, Rabi oscillations obtained via magnetic resonance ($B_2$-drive).
\textbf{c}, Stark shift in resonance frequency $\omega_e$ of the electron spin transition.
\textbf{d}, Dressed qubit Rabi oscillations obtained via electric resonance ($E_2$-drive).
\textbf{e}, Dressed qubit chevron Rabi pattern and \textbf{f}, spectrum obtained via frequency modulation resonance (FM-drive).
\textbf{g}, Level scheme of the dressed qubit system as a function of detuning $\Delta\omega=\omega_{\rm MW}-\omega_e$.
\textbf{h}, Dressed qubit Rabi frequency as a function of $\Delta\omega$ during the detuning pulse ($\Delta\omega$-drive).
}
\end{center}
\end{figure*}

All of our measurements are based on the initialization of the electron spin in the $\ket{\downarrow}$ state and readout of the electron spin $\ket{\uparrow}$ state via spin-dependent tunneling to the SET island~\cite{Morello2010}. If we want to work with the dressed qubit, we can continue using the same initialization and readout, however, we need to convert the electron $\ket{\downarrow}$, $\ket{\uparrow}$ states into the dressed qubit $\ket{-}$, $\ket{+}$ states and vice versa. This can be done in two ways. The first is to use resonant $\pi/2$-pulses along the $-y'$-axis in the frame of the MW source (see Fig.~\ref{figure03}\textbf{a}), which will convert $\ket{\downarrow}$ to $\ket{+}$ for initialization, and $\ket{+}$ to $\ket{\uparrow}$ for readout. Between initialization and readout the spin needs to be resonantly driven about the $x'$-axis to keep it spin-locked in the dressed picture~\cite{Abragam1961}. The second is to start with an off-resonant MW driving field along the $x'$-axis, and adiabatically reduce the detuning $\Delta\omega = \omega_{\rm MW} - \omega_e$ between the MW frequency $\omega_{\rm MW}$ and the ESR frequency $\omega_e$ to zero. This will convert $\ket{\downarrow}$ to $\ket{+}$ when reducing $\Delta\omega$ for initialization, and $\ket{-}$ to $\ket{\uparrow}$ when increasing $\Delta\omega$ for readout (also see Fig.~\ref{figure04}\textbf{g}).

\section{\label{control}Dressed qubit control} 
The Hamiltonian $H$ of the system is given in the basis of the spin states $\ket{\downarrow}$ and $\ket{\uparrow}$ by
\begin{equation}
\label{Hami}
H = \frac{1}{2} \hbar \gamma_e (B_0 \sigma_z + B_1 \cos(\omega_{MW} t) \sigma_x)
\end{equation}
in the lab frame, and
\begin{equation}
\label{Hamirot}
H_{\rm rot} = \frac{1}{2} \hbar (\Delta\omega \sigma_z + \Omega_R \sigma_x)
\end{equation}
in the rotating frame. As explained above, for a resonant coherent microwave drive the eigenstates are $\ket{+}=\frac{1}{\sqrt{2}}(\ket{\downarrow}+\ket{\uparrow})$ and $\ket{-}=\frac{1}{\sqrt{2}}(\ket{\downarrow}-\ket{\uparrow})$. We can change our basis to make these new eigenstates the basis states of our Hilbert space and we obtain the Hamiltonian in the dressed basis $H_\rho$, given by
\begin{equation}
\label{Hamirho}
H_\rho = \frac{1}{2} \hbar (\Omega_R \sigma_z+ \Delta\omega \sigma_x),
\end{equation}
where the quantization is determined by $\Omega_R$ and the coupling term by $\Delta\omega$. This offers a number of different methods to coherently control the dressed qubit. In the following we will introduce four different methods based on (i) magnetic resonance, modulating $\Delta\omega$ at frequency $\Omega_R$ via (ii) the Stark shift of $\omega_e$~\cite{Laucht2015} and via (iii) frequency modulation of $\omega_{\rm MW}$, and (iv) pulsing $\Delta\omega$ high by changing $\omega_{\rm MW}$ for a short period of time.

\subsection{\label{MagRes}Magnetic resonance}
Experimentally, the easiest method to control the dressed qubit is to apply a radiofrequency (RF) field $B_2$ with frequency $\omega_{\rm RF}=\Omega_R$ to the on-chip antenna. Here, only the magnetic field component perpendicular to the quantization axis of the dressed qubit drives the transition~\cite{Jeschke1999},\footnote{In principle another way of using magnetic resonace would be to use a second MW field at $\omega_e$ along $y'$ that is amplitude modulated at $\Omega_R$.} and the dressed qubit Rabi frequency is given by $\Omega_{R\rho} = \cos(\theta) \frac{1}{2} \hbar \gamma_e B_2$, where $\theta$ is the angle between the orientation of the external magnetic field $B_0$ and that of the oscillating RF field $B_2$. Simulations of the on-chip antenna~\cite{Dehollain2013} and the donor device~\cite{Laucht2015} indicate a non-zero $B_{2z'}$ magnitude at the donor location, with $\theta\approx 2^\circ$. 

We measure the spectral response of the dressed qubit by scanning $\omega_{\rm RF}$ over the frequency range around $\Omega_R$, and plot the results in Fig.~\ref{figure04}\textbf{a}. Here, the blue circles are the experimental result, while the red line is a fit to Rabi's formula. With $\omega_{\rm RF}=\Omega_R$ we can also measure coherent Rabi oscillations of the dressed qubit (see Fig.~\ref{figure04}\textbf{b}) with $\Omega_{R\rho} = 2.4$~kHz.

\subsection{\label{ElecRes}Electric resonance}
In the Hamiltonian in the dressed basis (Eq.~\ref{Hamirho}), the off-diagonal terms ($\sigma_x$ terms) directly depend on $\Delta\omega$. As we have shown recently~\cite{Laucht2015}, the electric field created by the gates can be used to induce a Stark shift of both the hyperfine coupling $A$ between the electron and the nucleus, and the gyromagnetic ratio $\gamma_e$ of the electron. In Fig.~\ref{figure04}\textbf{c}, we present the shift in resonance frequency $\omega_e$ for a DC voltage applied to the gates surrounding the donor~\footnote{For the measurements in this manuscript we Stark-shifted $\omega_e$ by applying a voltage to LDF and RDF (see Ref.~\cite{Laucht2015} for more information and comparison).}, and we extract a shift of $d\omega_e/dV=-167$~kHz/V. We can exploit the Stark shift to modulate $\Delta\omega$ at the dressed qubit's resonance frequency $\Omega_{R}$ to electrically drive Rabi oscillations of the dressed qubit. As this control method makes use of a local electric gate to modulate $\omega_e$, it is immediately compatible with scalable quantum computing architectures where the global driving field is kept unchanged during gate operations (see Sec.~\ref{Scale}).

For the measurement we connect the RF source directly to the gates to supply the oscillating electric field $E_2$. For an RF power of $P_{\rm RF}=+6$~dBm at the source, and $-30.5$~dB of attenuation, we obtain $P_{\rm RF}=-24.5$~dBm at the gates. This power translates to a voltage $V_{50\Omega}=18.8$~mV and the voltage applied to the gate is $V_{\rm gate}=2 V_{50\Omega} = 37.6$~mV. The shift in frequency is then $\Delta\omega_e=V_{\rm gate} (-167$~kHz/V$)=6.2$~kHz, which results in a dressed qubit Rabi frequency $\Omega_{R\rho}=\tfrac{1}{2} \Delta\omega_e=3.1$~kHz (due to oscillating vs. rotating field). We plot the experimentally obtained Rabi oscillations in Fig.~\ref{figure04}\textbf{d}. The black circles are the experimental data, and the red line is a fit to a sinusoid. $\Omega_{R\rho}$ is in excellent agreement with our prediction.

\subsection{\label{FMcont}FM control}
As an alternative to electrical modulation of $\Delta\omega$ via the Stark shift, we can also modulate $\Delta\omega$ by frequency modulation (FM) of $\omega_{\rm MW}$. The FM feature in our MW source provides a simpler way to experimentally implement this method, albeit with the lack of the scale-up potential provided by electric resonance (see Sec.~\ref{ElecRes}). This is because one has to keep track of the reference frequency defining the rotating frame, given by $\omega_{\rm MW}$. Another advantage of FM control is the much larger achievable Rabi frequency. The effective driving strength is only limited by the FM depth of the source and can in principle reach the regime where the driving strength exceeds the qubit splitting, $\Omega_{R\rho}>\omega_{\rm RF}$.

In Fig.~\ref{figure04}\textbf{e},\textbf{f} we plot the dressed qubit Rabi chevron pattern and spectrum (the red line is a fit to Rabi's formula), respectively. For these measurements, the output of an RF source is connected directly to the FM input of the MW source. The modulation amplitude $\Delta\omega_{\rm RF}$ can then be controlled by the power of the RF source and the FM bandwidth of the MW source, and $\Omega_{R\rho}=\tfrac{1}{2}\Delta\omega_{\rm RF}$.

\subsection{\label{DetPul}Detuning pulse}
The fourth and final method of dressed qubit control that we have investigated is realized by pulsing the detuning $\Delta\omega$ from 0 to a finite value for a short amount of time. For $\Delta\omega\gg\Omega_R$, the eigenstates are the spin basis states, and the dressed $\ket{+}$ state will nutate into the dressed $\ket{-}$ state performing a Rabi oscillation (see Fig.~\ref{figure04}\textbf{g}) with $\Omega_{R\rho}=\Delta\omega$. This control method is based on the same principle as the control of the state of the Cooper pair box~\cite{Nakamura1999} or the singlet-triplet qubit in semiconductor double quantum dots~\cite{DiVincenzo2000,Petta2005,Hanson2007a}. In principle, it is equivalent to pulse $\Delta\omega$ by tuning $\omega_e$ or $\omega_{\rm MW}$, albeit with the same limitations as for FM control (see Section~\ref{FMcont}). For our current sample design we can only reach $\Delta\omega\gg\Omega_R$ by pulsing $\omega_{\rm MW}$, which is why we limit our experiments to this case. With an  optimized gate layout we could also use local electric field control in a scalable manner.

In Fig.~\ref{figure04}\textbf{h}, we plot the Rabi frequency of the dressed qubit $\Omega_{R\rho}$ as a function of $\Delta\omega$ during the detuning pulse. For large detunings when $\Delta\omega \gg \Omega_R=460$~kHz the dressed qubit Rabi frequency follows $\Omega_{R\rho} = \Delta\omega$. This implies that the dressed qubit's Rabi frequency $\Omega_{R\rho}$ is larger than its level splitting $\Omega_R$. In fact, $\Omega_{R\rho} = 10$~MHz corresponds to gate operation times, one order of magnitude shorter than those normally obtained with pulsed ESR on the bare electron spin in our lab. However for smaller $\Delta\omega \sim \Omega_R$, $\Omega_{R\rho}$ tends towards $\Omega_R$ and the amplitude of the Rabi oscillations decreases. This behaviour is well described by Rabi's formula, and the red line in Fig.~\ref{figure04}\textbf{h} is simply calculated as $\Omega_{R\rho} = \sqrt{\Delta\omega^2 + \Omega_R^2}$. The raw data for this plot is presented in Fig.~\ref{figureS_DetPulse}.

\section{Dressed qubit lifetime and coherence times}

\begin{figure*}[!t]
\begin{center}
\includegraphics[width=0.5\textwidth]{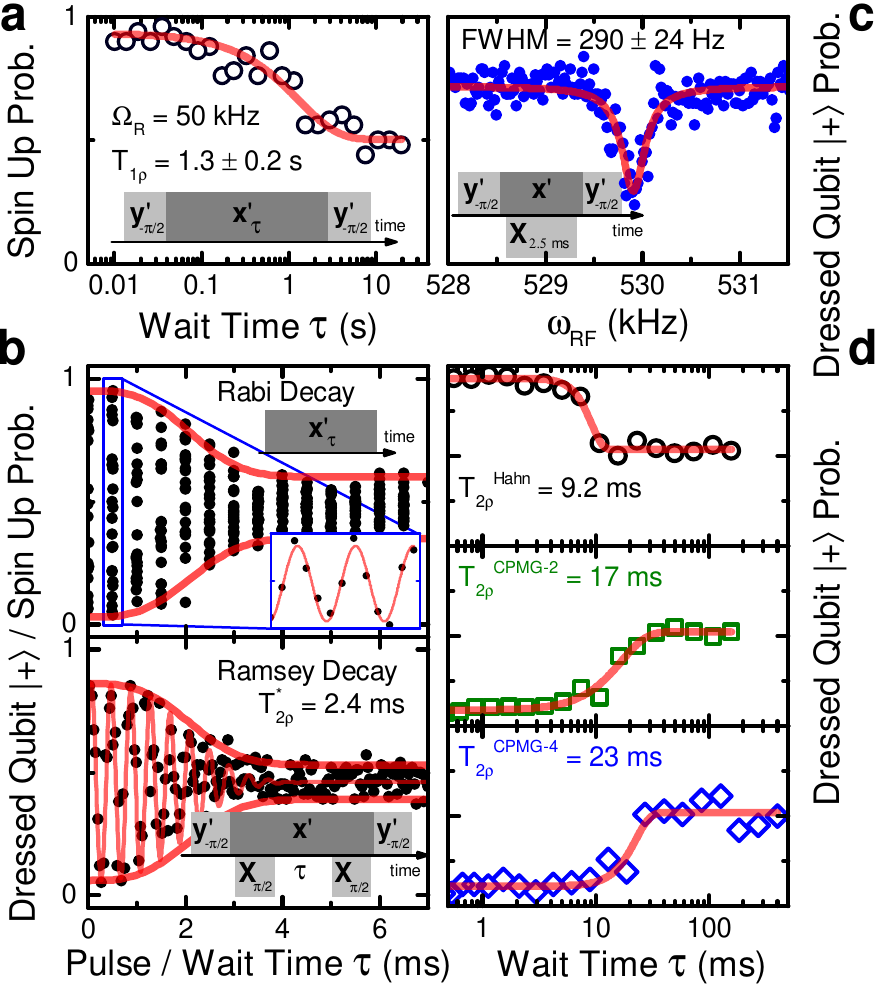}
\caption{\label{figure05} \textbf{Dressed qubit lifetime and coherence times.}
\textbf{a}, Longitudinal decay of the driven qubit $T_{1\rho}$. The inset shows the used pulse sequence.
\textbf{b}, Free induction decay $T_{2\rho}^{*}$ of the dressed qubit obtained from (upper panel) the decay of the electron spin Rabi oscillations measured at frequent intervals (the inset shows a zoom-in of the Rabi oscillations at around $\tau = 0.5$~ms) and (lower panel) a Ramsey sequence in the dressed frame (see insets for the pulse sequences).
\textbf{c}, Low RF power, dressed qubit spectrum showing the intrinsic linewidth (see inset for the pulse sequence).
\textbf{d}, Coherence times $T_{2\rho}^{\rm Hahn}$, $T_{2\rho}^{\rm CPMG-2}$ and $T_{2\rho}^{\rm CPMG-4}$ obtained from Hahn echos and CPMG sequences with 2 and 4 refocussing pulses, respectively.
}
\end{center}
\end{figure*}

We now take a look at the dressed qubit's lifetime $T_{1\rho}$ and coherence times $T_{2\rho}^*$ and $T_{2\rho}$~\footnote{The coherence time measurements in this section were performed with FM control.}. As the longitudinal decay time $T_{1\rho}$ in the driven frame is dependent on noise of frequency $\Omega_R$, it is often used for noise spectroscopy measurements (see Refs.~\cite{Ithier2005,Yan2013,Loretz2013} and section \ref{T1}). We perform $T_{1\rho}$ measurements for different Rabi frequencies $\Omega_R$. One example of a decay trace for $\Omega_R = 50$~kHz is shown in Fig.~\ref{figure05}\textbf{a}, with the inset showing the pulse sequence used. The decay curve yields $T_{1\rho}(50$~kHz$)=1.3$~s. For higher Rabi frequencies, we observe a decrease in $T_{1\rho}$ (see section~\ref{T1} and Fig.~\ref{figure03}\textbf{c}).

To obtain $T_{2\rho}^*$, we measure the free induction decay in two different ways. In Fig.~\ref{figure05}\textbf{b} (upper panel) we look at the decay of the Rabi oscillations of the driven electron spin. This is equivalent to the free precession in the driven frame. While the zoom-in shows the actual Rabi oscillations, we can estimate the decay of the envelope from the main plot. The red lines correspond to an exponential decay $\propto e^{-(\tau/T_{2\rho}^*)^3}$, with $T_{2\rho}^*=2.4$~ms. In Fig.~\ref{figure05}\textbf{b} (lower panel) we perform a Ramsey experiment in the driven frame. After the qubit is initialized in the $\ket{+}$ state, we use the standard pulse sequence $X_{\pi/2} - \tau - X_{\pi/2}$ (see inset) to measure the free induction decay. A slight detuning between $\omega_{\rm RF}$ and $\Omega_R$ leads to the Ramsey fringes, which can be fitted with a decaying sinusoid (red lines) with $T_{2\rho}^*=2.4$~ms. We also measure the low power transition linewidth (see Fig.~\ref{figure05}\textbf{c}) of the dressed spin and obtain a $\rm{FWHM}=290$~Hz for $P_{\rm RF}=-26$~dBm, close to the intrinsic limit given by $T_{2\rho}^*$.

The coherence time can be extended using dynamical decoupling and in Fig.~\ref{figure05}\textbf{d} we present the results of a Hahn echo, 2-pulse Carr-Purcell-Meiboom-Gill (CPMG), and a 4-pulse CPMG sequence~\cite{Hahn1950,Meiboom1958}. We obtain $T_{2\rho}^{\rm Hahn}=9.2$~ms, $T_{2\rho}^{\rm CPMG-2}=17$~ms, and $T_{2\rho}^{\rm CPMG-4}=23$~ms, respectively. Overall, the coherence times are an order of magnitude longer than those of the undriven spin~\cite{Muhonen2014}, and are in line with similar measurements on microwave-dressed atomic ions~\cite{Timoney2011} and NV centres in diamond~\cite{Rabl2009,Xu2012,Golter2014}. The improvement of the coherence times is achieved by the intrinsic insensitivity of the dressed states to magnetic field fluctuations. Compared with dynamical decoupling sequences, where effects of a fluctuating background are refocussed at specific times, the dressed spin state provides continuous protection from decoherence.

\section{\label{Scale}Scalability}
The use of the dressed spin states as qubit states provides some interesting advantages for scaled-up quantum computation architectures~\cite{Mikelsons2015,Cai2015}. Two of the four dressed qubit control mechanisms introduced in Section~\ref{control} are compatible with scalable architectures based on a global, always-on microwave field and local gate operations~\cite{Kane1998,Laucht2015}. This is because the dressed qubit can be controlled using local electric gates that apply an oscillating electric field (see Section~\ref{ElecRes}) or electric field pulses (see Section~\ref{DetPul}). The electric field Stark shifts the gyromagnetic ratio and the hyperfine coupling, which are terms occupying the off-diagonal elements in the driven qubit Hamiltonian (Equ.~\ref{Hamirho}).

Coupling of two dressed qubits and the realization of two-qubit logic gates can be realized by spin-spin coupling via Hartmann-Hahn double resonance~\cite{Hartmann1962,Cai2013,London2013}. Considering two spins with slightly different hyperfine couplings due to Stark or strain shifts, both can be dressed with microwave fields resonant with their transition frequencies $\omega_{e1} \neq \omega_{e2}$, creating dressed qubits with level splittings $\Omega_{R1}$ and $\Omega_{R2}$, respectively. By choosing the two driving powers appropriately, the two dressed qubits can be tuned in resonance with each other ($\Omega_{R1}=\Omega_{R2}$), and coupled via exchange coupling or magnetic dipolar coupling~\cite{Hartmann1962}. Although smaller than exchange coupling for small distances, magnetic dipolar coupling decays more slowly, making it the dominant coupling mechanism for distances exceeding a few tens of nm. At a donor separation of $50$~nm, a coupling strength of up to $400$~Hz is achievable, allowing a number of gate operations to be conducted within the coherence time of the dressed qubit, assuming dynamical decoupling. Furthermore, if a pair of quantum systems can be dressed by the same driving field, a hybrid dressed state can be created that is insensitive to both amplitude and phase noise in the continuous driving field~\cite{Cai2015}.

Another interesting property that is enabled in the dressed basis is coupling of the spin states to strain in the silicon crystal lattice~\cite{Dreher2011}, which opens the possibility to coherent qubit control by mechanical means~\cite{Franke2015,Barfuss2015}, and coherent spin-phonon dynamics~\cite{Soykal2011,Gustafsson2014}. One could even envision coupling the dressed spins to nanomechanical oscillators~\cite{Rabl2009}, and realize long-distance spin-spin coupling via the quantized modes of the oscillator. For this coupling scheme, a first MW field at $\omega_{\rm MW1}$ would supply the dressing of the qubit with level splitting $\Omega_{R1}$, while a local electric gate would control the dressed single qubit rotations via an oscillating electric field. A DC electric field could then be used to Stark-shift the spin transition into resonance with a second MW field at $\omega_{\rm MW2}\neq\omega_{\rm MW1}$ with strength $\Omega_{R2}\neq\Omega_{R1}$. This second MW drive would enable coupling to the nanomechanical oscillator, by bringing the dressed qubit into resonance with the oscillator's frequency at $\omega_{r}=\Omega_{R2}$.

\section{Summary \& conclusions}
In conclusion, our work demonstrates the use of the dressed states of an electron spin on a phosphorus donor in silicon as a qubit. We first prove the existence of dressed states by observing Mollow triplets in a pump-probe experiment and performing Rabi oscillations. We demonstrate 4 different control methods of the dressed qubit, of which two are compatible with scalable quantum computing architectures, one results in faster gate operations than those obtainable with pulsed spin resonance on the bare electron spin, and one achieves Rabi frequencies larger than the level splitting of the system. This variety of control methods is available due the dressed basis, which also unlocks coupling to electric fields and lattice strain, opening promising avenues for future research and applications. Furthermore, we find coherence times that are one order of magnitude longer that those of the bare spin, with $T_{2\rho}^*=2.4$~ms and $T_{2\rho}^{\rm Hahn}=9.2$~ms. Overall, this work provides a pathway to apply to a solid-state platform some advanced control methods traditionally reserved to optical and atomic systems, greatly expanding the potential of donor spin qubits for applications in quantum information processing and nanoscale research.

\bibliography{Papers}

\begin{addendum}
 \item[Acknowledgements] This research was funded by the Australian Research Council Centre of Excellence for Quantum Computation and Communication Technology (project number CE110001027) and the US Army Research Office (W911NF-13-1-0024). We acknowledge support from the Australian National Fabrication Facility, and from the laboratory of Prof. Robert Elliman at the Australian National University for the ion implantation facilities. The work at Keio has been supported by JSPS KAKEN (S) and Core-to-Core Program.
 \item[Authors Contributions] A.L., R.K., S.S., J.P.D., J.T.M., A.S.D. and A.M. designed the experiments. A.L. performed the measurements and analysed the results with A.M.'s supervision and R.K.'s and S.S.'s assistance. A.L. and F.A.M. performed the simulations with A.M.'s supervision. D.N.J. and J.C.M. designed and performed the $P$ implantation experiments. F.E.H. fabricated the device with A.S.D.'s supervision and R.K.'s and S.F.'s assistance. K.M.I. prepared and supplied the $^{28}$Si epilayer wafer. A.L. and A.M. wrote the manuscript, with input from all coauthors.
 \item[Competing financial interests] The authors declare that they have no competing financial interests.
 \item[Correspondence] Correspondence and requests for materials should be addressed to A.L.~(a.laucht@unsw.edu.au) or A.M.~(a.morello@unsw.edu.au).
\item[Supplementary Information] accompanies the paper.
\end{addendum}

\end{multicols}
\newpage
\setlength{\columnsep}{1cm}
\begin{multicols}{2}
\setstretch{1.15}

\bibliographystyle{naturemag}

\renewcommand\thesection{}
\section{Supplementary Information}

\renewcommand\thesubsection{S\arabic{subsection}}

\setcounter{page}{1}
\renewcommand{\thepage}{S\arabic{page}}

\renewcommand\thesuppfig{S\arabic{suppfig}}

\subsection{Device Fabrication}
The device was fabricated on a $0.9$~$\mu$m thick epilayer of isotopically purified $^{28}$Si ($800$~ppm residual $^{29}$Si concentration), grown on top of a $500$~$\mu$m thick $^{\rm nat}$Si wafer~\cite{Itoh2014}. Single-atom qubits were selected out of a small group of donors implanted in a region adjacent to the Single-Electron-Transistor (SET). In this device, P$^+_2$ molecular ions were implanted at $20$~keV energy in a $100\times100$~nm$^2$ window.
All other nanofabrication processes were identical to those described in detail in Ref.~\cite{Pla2012}, except for a slight modification in the gate layout to bring the qubits closer to the microwave antenna and provide an expected factor $3\times$ improvement in $B_1$ (see Fig.~\ref{figure01}\textbf{a} and Ref.~\cite{Laucht2015} for exact gate layout).

\subsection{Experimental Setup}
The sample was mounted on a high-frequency printed circuit board in a copper enclosure, thermally anchored to the cold finger of an Oxford Kelvinox $100$ dilution refrigerator with a base temperature $T_{\rm bath}=20$~mK. The sample was placed in the center of a wide-bore superconducting magnet, oriented so that the $B_0$ field was applied in the plane of the silicon chip, along the [110] crystal axis, and perpendicular to the short-circuit termination of the MW antenna. The magnet was operated in persistent mode while also feeding the nominal current through the external leads. We found that removing the supply current while in persistent mode led to a very significant magnetic field and ESR frequency drift, unacceptable given the intrinsic sharpness of the resonance lines of our qubit. Conversely, opening the persistent mode switch led to noticeable deterioration of the spin coherence, most visible as a shortening of $T^\ast_2$ in Ramsey experiments.

Room-temperature voltage noise was filtered using an anti-inductively wound coil of thin copper wire with a core of Eccosorb CRS-117 ($\sim1$~GHz cut-off), followed by two types of passive low-pass filters: $200$~Hz second-order $RC$ filters for DC biased lines, and $80$~MHz seventh-order Mini-Circuits LC filters for pulsed voltage lines. The filter assemblies were placed in copper enclosures, filled with copper powder, and thermally anchored to the mixing chamber. DC voltages were applied using optoisolated and battery-powered voltage sources, connected to the cold filter box via twisted-pair wires. Voltage pulses were applied using an arbitrary waveform generator (LeCroy ArbStudio $1104$), connected to the filter box via semi-rigid coaxial lines. MW excitations were generated using either an Agilent E8257D analog or an Agilent E8267D vector signal generator, and RF excitations were produced by an Agilent MXG N5182A vector signal generator. Both excitation signals were combined using a power-combiner and fed to the MW antenna via a CuNi semi-rigid coaxial cable, with attenuators at the $1.5$~K stage ($10$~dB) and the $20$~mK stage (3 dB). The SET current was measured by a Femto DLPCA-200 transimpedance amplifier at room temperature, followed by a floating-input voltage post-amplifier, a sixth-order low-pass Bessel filter, and acquired using a PCI digitiser card (AlazarTech ATS9440). For electron spin experiments the state is always initialized spin-down and all of our plots were produced by taking the spin-up proportion from $100-200$ single-shot measurement repetitions per point. 

\begin{suppfig*}[!b]
\begin{center}
\includegraphics[width=0.55\textwidth]{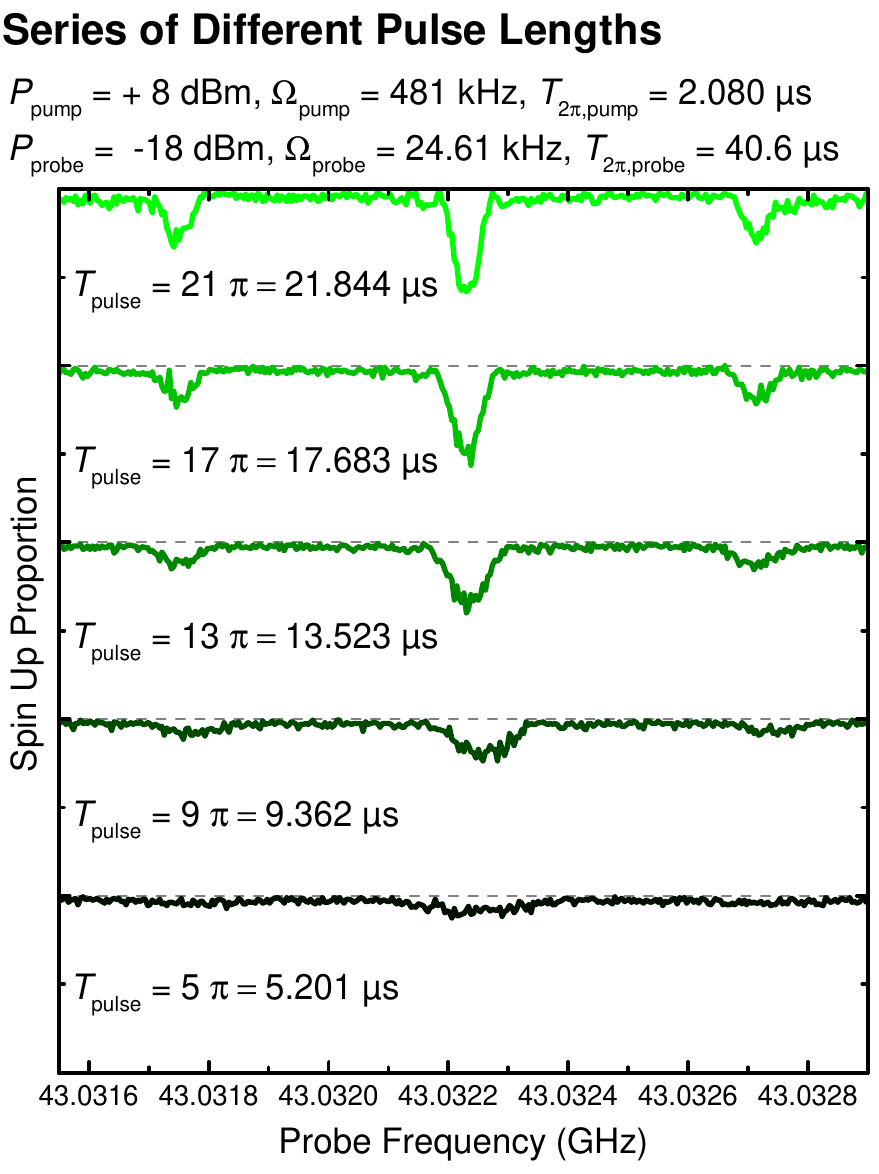}
\caption{\label{figureS_pulselength} \textbf{Dressing the electron spin.} Mollow spectra of the dressed electron spin for different microwave pulse lengths. Here, a strong, resonant driving field $P_{\rm pump}$ was used to dress the spin state, while a weaker probe field $P_{\rm probe} = P_{\rm pump}-26$~dB was scanned over the Mollow triplet to record the spectra.
}
\end{center}
\end{suppfig*}

\newpage

\subsection{\label{SuppMollow}Mollow Triplet Measurements}

\begin{suppfig*}[!t]
\begin{center}
\includegraphics[width=0.55\textwidth]{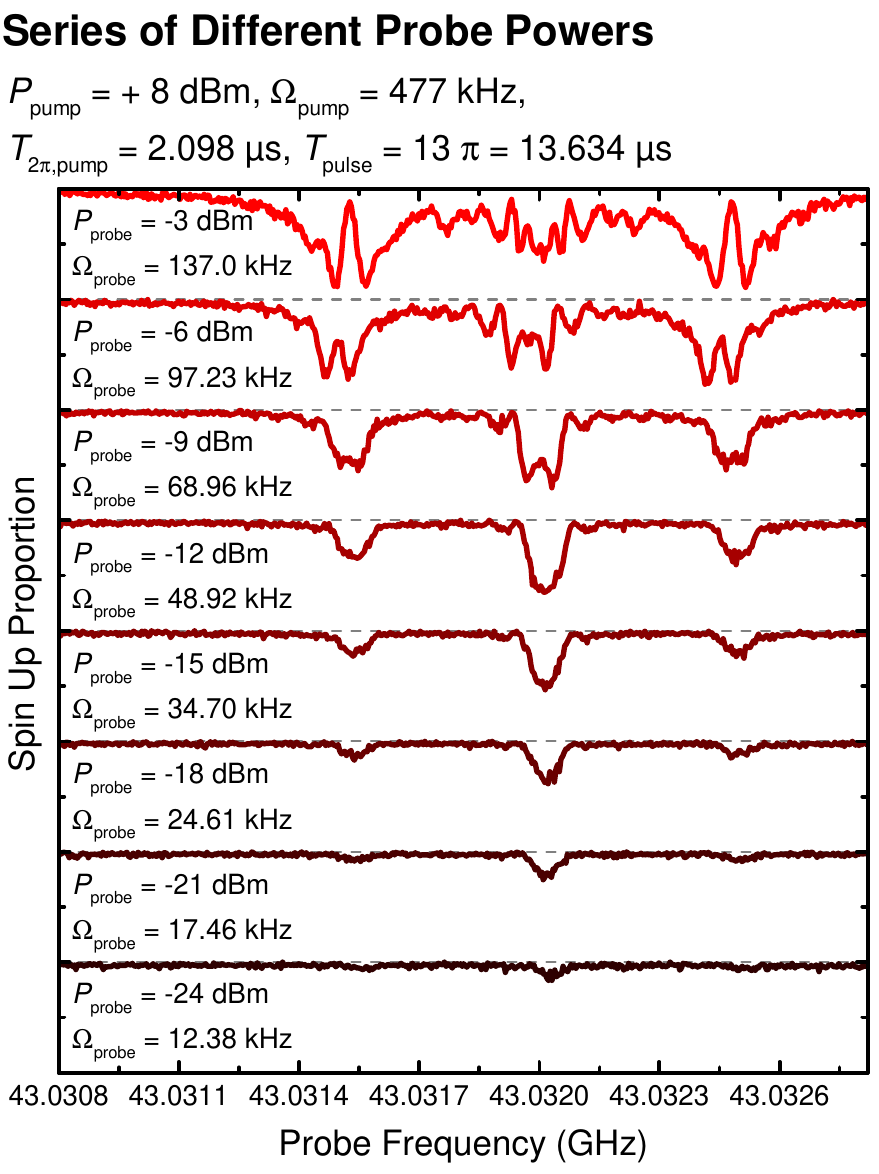}
\caption{\label{figureS_probepower} \textbf{Dressing the electron spin.} Mollow spectra of the dressed electron spin for different probe powers.
}
\end{center}
\end{suppfig*}

In addition to the data presented in the main text (Fig.~\ref{figure02}), we recorded some spectra where the microwave pulse length was varied (Fig.~\ref{figureS_pulselength}) and where the probe power was varied (Fig.~\ref{figureS_probepower}). 

Fig.~\ref{figureS_pulselength} shows the Mollow spectrum for different pulse lengths from $T_{\rm pulse}=5\pi$ at the bottom to $T_{\rm pulse}=21\pi$ at the top. Here, $P_{\rm pump}=+8$~dBm and $P_{\rm probe}=-18$~dBm were kept constant for all measurements. The pump pulse always rotates the electron spin spin by $(2\mathrm{n}+1)\cdot\pi$ from the $\ket{\downarrow}$ to the $\ket{\uparrow}$ state for all of the chosen pulse lengths. The simultaneous probe pulse at $P_{\rm probe} = P_{\rm pump}-26$~dB rotates the electron by $\approx\pi$ for $T_{\rm pulse}=21\pi$, for which the Mollow triplet can be clearly seen. For shorter pulse lengths, the probe pulse has a smaller effect on the electron spin, and produces only small dips in the spectrum.

\newpage
Fig.~\ref{figureS_probepower} shows the dependence of the Mollow spectrum for probe powers from $P_{\rm probe}=-24$~dBm at the bottom to $P_{\rm probe}=-3$~dBm at the top. Here, $P_{\rm pump}=+8$~dBm and $T_{\rm pulse}=13\pi$ were kept constant for all measurements. For low $P_{\rm probe}$ the probe pulse only has little influence on the electron spin. For $P_{\rm probe}=-18$~dBm, we recover the case for Fig.~\ref{figureS_pulselength}, and for $P_{\rm probe}>-12$~dBm, the probe pulse actually overrotates the electron spin and the Mollow spectrum shows a multipeak signature, similar to the normal Rabi spectrum.

\clearpage

\subsection{\label{DetPulse}Dressed Qubit Control by Detuning Pulse}

One of the four methods to control the dressed qubit is realized by pulsing the detuning $\Delta\omega = \omega_{\rm MW} - \omega_e$ to a finite value for a short amount of time before $\Delta\omega$ is pulsed back to zero. This means that the driving field and the electron spin will rotate at different frequencies $\omega_{\rm MW} \neq \omega_e$, and accumulate a phase term (like a $z$-gate). This constitutes a qubit rotation in the driven frame. In Fig.~\ref{figure04}\textbf{h} we have shown that the Rabi frequency of the dressed qubit $\Omega_{R\rho}$ is always larger than $\Omega_R=460$~kHz, albeit with a decrease in the amplitude of the Rabi oscillations when $\Delta\omega \lesssim \Omega_R$. We plot the supporting raw data for this plot in Fig.~\ref{figureS_DetPulse}. The panels show the Rabi oscillations obtained from $\Delta\omega=0.25$~MHz~$<\Omega_R$ at the bottom, to $\Delta\omega=2.8$~MHz~$>\Omega_R$ at the top. The frequency of the oscillations follows the formula $\Omega_{R\rho} = \sqrt{\Delta\omega^2 + \Omega_R^2}$ (also see section~\ref{DetPul}) and is limited to $\Omega_{R\rho}\geq\Omega_R$. Since the amplitude of the oscillations is given by $A_{R\rho}=\frac{\Delta\omega^2}{\Delta\omega^2+\Omega_R^2}$, high-fidelity operations require $\Delta\omega\gg\Omega_R$.

\end{multicols}
\renewcommand{\thepage}{S\arabic{page}}
\renewcommand\thesuppfig{S\arabic{suppfig}}
\begin{suppfig}[!b]
\begin{center}
\includegraphics[width=0.55\textwidth]{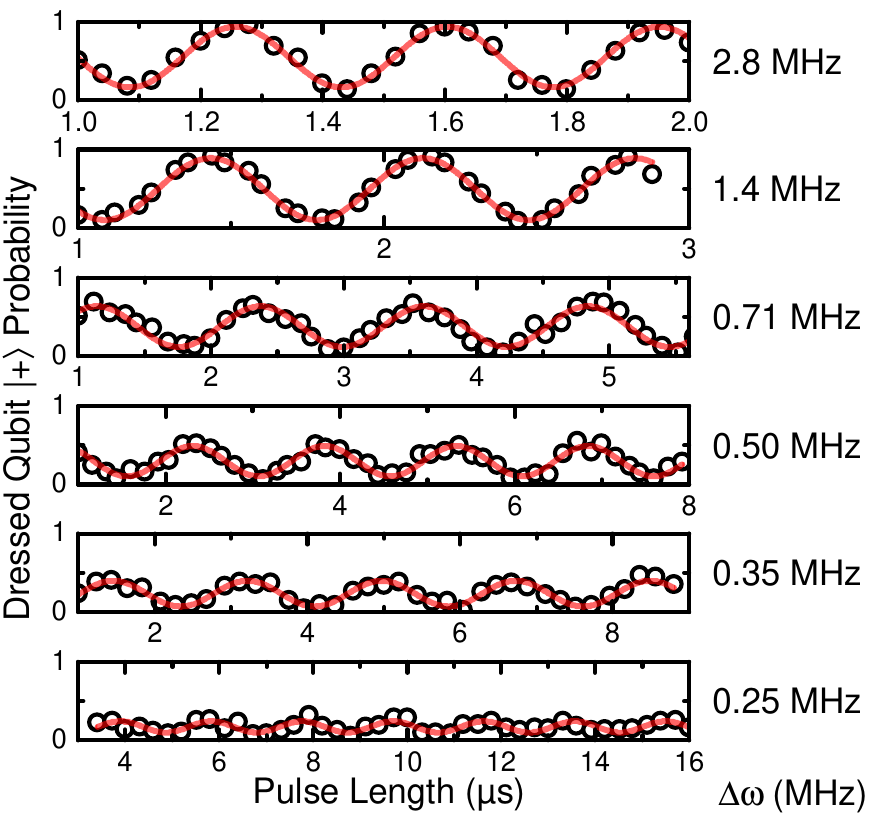}
\caption{\label{figureS_DetPulse} \textbf{Dressed qubit control by detuning pulse.} Rabi oscillations of the dressed qubit for different detuning amplitudes $\Delta\omega$ during the detuning pulse. Once $\Delta\omega\leq\Omega_R$ the amplitude of the Rabi oscillations decreases. See also Sec.~\ref{DetPul}.
}
\end{center}
\end{suppfig}

\clearpage
\begin{multicols}{2}
\setstretch{1.15}
\renewcommand\thesubsection{S\arabic{subsection}}
\renewcommand\thesuppfig{S\arabic{suppfig}}
\renewcommand{\thepage}{S\arabic{page}}

\subsection{\label{T1}Dressed qubit lifetime and noise spectroscopy in the driven frame} 

\begin{suppfig*}[!t]
\begin{center}
\includegraphics[width=1\columnwidth]{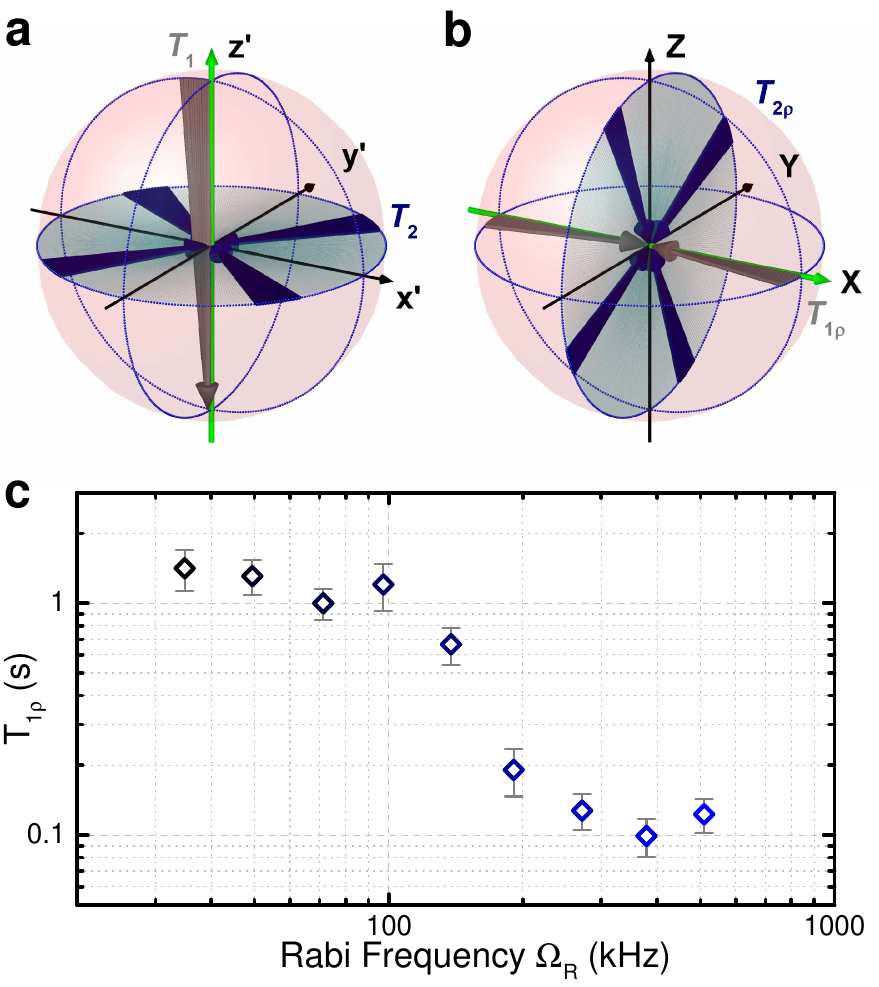}
\caption{\label{figure03} \textbf{Noise spectroscopy in the rotating frame.}
\textbf{a}, Free evolution and decay in the rotating frame, and 
\textbf{b}, driven evolution and decay in the driven, double rotating frame. The green axes indicates the quantization axes. The temperature-dependent longitudinal ($T_1$ and $T_{1\rho}$) depolarization is indicated by grey arrows and the transverse ($T_2$ and $T_{2\rho}$) depolarization by blue arrows.
\textbf{c}, Longitudinal decay of the driven qubit $T_{1\rho}$ as a function of the Rabi frequency $\Omega_R$.
}
\end{center}
\end{suppfig*}

The dressed spin state offers an interesting method to perform noise spectroscopy. As the longitudinal decay $T_{1\rho}$ is dependent on the noise that is perpendicular to the quantization axis of the quantum state and resonant with the level splitting, $T_{1\rho}$ depends on $\Omega_R$. $T_{1\rho}(\Omega_R)$ reflects the spectrum of the magnetic noise in Z-direction, as any noise in Y-direction at $\Omega_R$ will get averaged out~\cite{Ithier2005,Yan2013}. This means that the dressed qubit can be used as highly sensitive probe for magnetic fields oscillating at frequency $\Omega_R$~\cite{Loretz2013}, and for the detection of individual nuclear spins~\cite{Mkhitaryan2015}.

In Fig.~\ref{figure03}\textbf{a},\textbf{b} we compare the spin depolarization dynamics in the rotating frame and the driven frame, respectively. In the rotating frame (Fig.~\ref{figure03}\textbf{a}), the quantization axis of the qubit (indicated in green) is oriented along the $z'$-axis, and longitudinal decay $T_1$ occurs from the north pole to the south pole of the Bloch sphere (grey arrow). During free precession the spin would be static in the rotating frame, pointing somewhere along the $x'-y'$-plane (shaded in cyan) and transverse depolarization $T_2$ would occur from the equator towards the centre of the Bloch sphere (dark blue arrows)~\footnote{The centre of the Bloch sphere represents a completely mixed state.}. In the driven frame (Fig.~\ref{figure03}\textbf{b}), we assume a driving field along the $X$-axis which constitutes the quantization axis here (indicated in green). Longitudinal decay $T_{1\rho}$ now occurs from $\pm X$ to the centre of the Bloch sphere (grey arrows), as the $\ket{+}$-$\ket{-}$ level splitting is much smaller than the thermal energy at $\sim100$~mK. During free precession the state is static in the double rotating frame (i.e. the reference frame that rotates around $B_0$ and $B_1$), pointing somewhere along the $Y-Z$-plane (shaded in cyan). Transverse depolarization $T_{2\rho}$ then occurs from there to the centre of the Bloch sphere (dark blue arrows).

We perform $T_{1\rho}$ measurements in the driven frame for different Rabi frequencies $\Omega_R$. Varying $\Omega_R$ allows us to map out $T_{1\rho}(\Omega_R)$ (see Fig.~\ref{figure03}\textbf{c}). This method offers access to the noise spectrum in a higher frequency range as compared to the standard spectroscopy via dynamical decoupling pulses~\cite{Muhonen2014}, which was limited to frequencies below $50$~kHz, dictated by the inverse of the minimum interval between pulses. For $\Omega_R<100$~kHz, we measure $T_{1\rho}>1$~s, but for larger $\Omega_R$ this value reduces to $T_{1\rho}\sim120$~ms. Earlier experiments on the noise spectrum measured by dynamical decoupling~\cite{Muhonen2014} showed white noise between $10$and $50$~kHz. This was attributed to broadband thermal noise that would not give rise to an increase in noise amplitude at higher frequencies. An explanation for the apparent increase in noise power above $100$~kHz in the $T_{1\rho}$ measurement could be the increasing influence of amplitude fluctuations of $B_1$ when $\Omega_R$ is increased. This is because $B_1$ has a component $B_{1z}$ parallel to the external magnetic field $B_0$.

\bibliography{Papers}

\begin{thebibliography}{10}
\expandafter\ifx\csname url\endcsname\relax
  \def\url#1{\texttt{#1}}\fi
\expandafter\ifx\csname urlprefix\endcsname\relax\def\urlprefix{URL }\fi
\providecommand{\bibinfo}[2]{#2}
\providecommand{\eprint}[2][]{\url{#2}}

\bibitem{Mollow1969}
\bibinfo{author}{Mollow, B.~R.}
\newblock \bibinfo{title}{Power spectrum of light scattered by two-level
  systems}.
\newblock \emph{\bibinfo{journal}{Phys. Rev.}} \textbf{\bibinfo{volume}{188}},
  \bibinfo{pages}{1969--1975} (\bibinfo{year}{1969}).

\bibitem{Xu2007}
\bibinfo{author}{Xu, X.} \emph{et~al.}
\newblock \bibinfo{title}{Coherent optical spectroscopy of a strongly driven
  quantum dot}.
\newblock \emph{\bibinfo{journal}{Science}} \textbf{\bibinfo{volume}{317}},
  \bibinfo{pages}{929--932} (\bibinfo{year}{2007}).

\bibitem{Baur2009}
\bibinfo{author}{Baur, M.} \emph{et~al.}
\newblock \bibinfo{title}{Measurement of autler-townes and mollow transitions
  in a strongly driven superconducting qubit}.
\newblock \emph{\bibinfo{journal}{Phys. Rev. Lett.}}
  \textbf{\bibinfo{volume}{102}}, \bibinfo{pages}{243602}
  (\bibinfo{year}{2009}).
\newblock
  \urlprefix\url{http://link.aps.org/doi/10.1103/PhysRevLett.102.243602}.

\bibitem{London2013}
\bibinfo{author}{London, P.} \emph{et~al.}
\newblock \bibinfo{title}{Detecting and polarizing nuclear spins with double
  resonance on a single electron spin}.
\newblock \emph{\bibinfo{journal}{Physical review letters}}
  \textbf{\bibinfo{volume}{111}}, \bibinfo{pages}{067601}
  (\bibinfo{year}{2013}).

\bibitem{Hartmann1962}
\bibinfo{author}{Hartmann, S.~R.} \& \bibinfo{author}{Hahn, E.~L.}
\newblock \bibinfo{title}{Nuclear double resonance in the rotating frame}.
\newblock \emph{\bibinfo{journal}{Phys. Rev.}} \textbf{\bibinfo{volume}{128}},
  \bibinfo{pages}{2042--2053} (\bibinfo{year}{1962}).
\newblock \urlprefix\url{http://link.aps.org/doi/10.1103/PhysRev.128.2042}.

\bibitem{Cai2013}
\bibinfo{author}{Cai, J.}, \bibinfo{author}{Jelezko, F.},
  \bibinfo{author}{Plenio, M.~B.} \& \bibinfo{author}{Retzker, A.}
\newblock \bibinfo{title}{Diamond-based single-molecule magnetic resonance
  spectroscopy}.
\newblock \emph{\bibinfo{journal}{New Journal of Physics}}
  \textbf{\bibinfo{volume}{15}}, \bibinfo{pages}{013020}
  (\bibinfo{year}{2013}).
\newblock \urlprefix\url{http://stacks.iop.org/1367-2630/15/i=1/a=013020}.

\bibitem{Timoney2011}
\bibinfo{author}{Timoney, N.} \emph{et~al.}
\newblock \bibinfo{title}{Quantum gates and memory using microwave-dressed
  states}.
\newblock \emph{\bibinfo{journal}{Nature}} \textbf{\bibinfo{volume}{476}},
  \bibinfo{pages}{185--188} (\bibinfo{year}{2011}).

\bibitem{Laucht2015}
\bibinfo{author}{Laucht, A.} \emph{et~al.}
\newblock \bibinfo{title}{Electrically controlling single-spin qubits in a
  continuous microwave field}.
\newblock \emph{\bibinfo{journal}{Science Advances}}
  \textbf{\bibinfo{volume}{1}}, \bibinfo{pages}{1500022}
  (\bibinfo{year}{2015}).

\bibitem{Dreher2011}
\bibinfo{author}{Dreher, L.} \emph{et~al.}
\newblock \bibinfo{title}{Electroelastic hyperfine tuning of phosphorus donors
  in silicon}.
\newblock \emph{\bibinfo{journal}{Phys. Rev. Lett.}}
  \textbf{\bibinfo{volume}{106}}, \bibinfo{pages}{037601}
  (\bibinfo{year}{2011}).

\bibitem{Rabl2009}
\bibinfo{author}{Rabl, P.} \emph{et~al.}
\newblock \bibinfo{title}{Strong magnetic coupling between an electronic spin
  qubit and a mechanical resonator}.
\newblock \emph{\bibinfo{journal}{Physical Review B}}
  \textbf{\bibinfo{volume}{79}}, \bibinfo{pages}{041302}
  (\bibinfo{year}{2009}).

\bibitem{Mikelsons2015}
\bibinfo{author}{Mikelsons, G.}, \bibinfo{author}{Cohen, I.},
  \bibinfo{author}{Retzker, A.} \& \bibinfo{author}{Plenio, M.~B.}
\newblock \bibinfo{title}{Universal set of gates for microwave dressed-state
  quantum computing}.
\newblock \emph{\bibinfo{journal}{New Journal of Physics}}
  \textbf{\bibinfo{volume}{17}}, \bibinfo{pages}{053032}
  (\bibinfo{year}{2015}).
\newblock \urlprefix\url{http://stacks.iop.org/1367-2630/17/i=5/a=053032}.

\bibitem{Cai2015}
\bibinfo{author}{Cai, J.}, \bibinfo{author}{Cohen, I.},
  \bibinfo{author}{Retzker, A.} \& \bibinfo{author}{Plenio, M.~B.}
\newblock \bibinfo{title}{Proposal for high-fidelity quantum simulation using a
  hybrid dressed state}.
\newblock \emph{\bibinfo{journal}{Phys. Rev. Lett.}}
  \textbf{\bibinfo{volume}{115}}, \bibinfo{pages}{160504}
  (\bibinfo{year}{2015}).
\newblock
  \urlprefix\url{http://link.aps.org/doi/10.1103/PhysRevLett.115.160504}.

\bibitem{Pla2012}
\bibinfo{author}{Pla, J.~J.} \emph{et~al.}
\newblock \bibinfo{title}{A single-atom electron spin qubit in silicon}.
\newblock \emph{\bibinfo{journal}{Nature (London)}}
  \textbf{\bibinfo{volume}{489}}, \bibinfo{pages}{541--545}
  (\bibinfo{year}{2012}).

\bibitem{Itoh2014}
\bibinfo{author}{Itoh, K.~M.} \& \bibinfo{author}{Watanabe, H.}
\newblock \bibinfo{title}{Isotope engineering of silicon and diamond for
  quantum computing and sensing applications}.
\newblock \emph{\bibinfo{journal}{MRS Communications}}
  \textbf{\bibinfo{volume}{4}}, \bibinfo{pages}{143--157}
  (\bibinfo{year}{2014}).
\newblock
  \urlprefix\url{http://journals.cambridge.org/article_S2159685914000329}.

\bibitem{Jamieson2005}
\bibinfo{author}{Jamieson, D.~N.} \emph{et~al.}
\newblock \bibinfo{title}{{Controlled shallow single-ion implantation in
  silicon using an active substrate for sub-20-keV ions}}.
\newblock \emph{\bibinfo{journal}{Applied Physics Letters}}
  \textbf{\bibinfo{volume}{86}}, \bibinfo{pages}{202101}
  (\bibinfo{year}{2005}).
\newblock
  \urlprefix\url{http://link.aip.org/link/APPLAB/v86/i20/p202101/s1\&Agg=doi}.

\bibitem{Morello2009}
\bibinfo{author}{Morello, A.} \emph{et~al.}
\newblock \bibinfo{title}{{Architecture for high-sensitivity single-shot
  readout and control of the electron spin of individual donors in silicon}}.
\newblock \emph{\bibinfo{journal}{Phys. Rev. B}} \textbf{\bibinfo{volume}{80}},
  \bibinfo{pages}{081307(R)} (\bibinfo{year}{2009}).
\newblock \urlprefix\url{http://link.aps.org/doi/10.1103/PhysRevB.80.081307}.

\bibitem{Morello2010}
\bibinfo{author}{Morello, A.} \emph{et~al.}
\newblock \bibinfo{title}{Single-shot readout of an electron spin in silicon}.
\newblock \emph{\bibinfo{journal}{Nature (London)}}
  \textbf{\bibinfo{volume}{467}}, \bibinfo{pages}{687--691}
  (\bibinfo{year}{2010}).

\bibitem{Dehollain2013}
\bibinfo{author}{Dehollain, J.~P.} \emph{et~al.}
\newblock \bibinfo{title}{Nanoscale broadband transmission lines for spin qubit
  control}.
\newblock \emph{\bibinfo{journal}{Nanotechnology}}
  \textbf{\bibinfo{volume}{24}}, \bibinfo{pages}{015202}
  (\bibinfo{year}{2013}).

\bibitem{Abragam1961}
\bibinfo{author}{Abragam, A.}
\newblock \emph{\bibinfo{title}{The principles of nuclear magnetism}}.
\newblock \bibinfo{number}{32} (\bibinfo{publisher}{Oxford university press},
  \bibinfo{year}{1961}).

\bibitem{Vandersypen2005}
\bibinfo{author}{Vandersypen, L.} \& \bibinfo{author}{Chuang, I.}
\newblock \bibinfo{title}{{NMR techniques for quantum control and
  computation}}.
\newblock \emph{\bibinfo{journal}{Reviews of Modern Physics}}
  \textbf{\bibinfo{volume}{76}}, \bibinfo{pages}{1037--1069}
  (\bibinfo{year}{2005}).
\newblock \urlprefix\url{http://link.aps.org/doi/10.1103/RevModPhys.76.1037}.

\bibitem{Jelezko2004}
\bibinfo{author}{Jelezko, F.} \emph{et~al.}
\newblock \bibinfo{title}{Observation of coherent oscillation of a single
  nuclear spin and realization of a two-qubit conditional quantum gate}.
\newblock \emph{\bibinfo{journal}{Phys. Rev. Lett.}}
  \textbf{\bibinfo{volume}{93}}, \bibinfo{pages}{130501}
  (\bibinfo{year}{2004}).

\bibitem{Koppens2006}
\bibinfo{author}{Koppens, F. H.~L.} \emph{et~al.}
\newblock \bibinfo{title}{{Driven coherent oscillations of a single electron
  spin in a quantum dot.}}
\newblock \emph{\bibinfo{journal}{Nature}} \textbf{\bibinfo{volume}{442}},
  \bibinfo{pages}{766--71} (\bibinfo{year}{2006}).
\newblock \urlprefix\url{http://www.ncbi.nlm.nih.gov/pubmed/16915280}.

\bibitem{Press2008}
\bibinfo{author}{Press, D.}, \bibinfo{author}{Ladd, T.~D.},
  \bibinfo{author}{Zhang, B.} \& \bibinfo{author}{Yamamoto, Y.}
\newblock \bibinfo{title}{Complete quantum control of a single quantum dot spin
  using ultrafast optical pulses}.
\newblock \emph{\bibinfo{journal}{Nature}} \textbf{\bibinfo{volume}{456}},
  \bibinfo{pages}{218--221} (\bibinfo{year}{2008}).

\bibitem{Muhonen2014}
\bibinfo{author}{Muhonen, J.~T.} \emph{et~al.}
\newblock \bibinfo{title}{Storing quantum information for 30 seconds in a
  nanoelectronic device}.
\newblock \emph{\bibinfo{journal}{Nature Nanotechnology}}
  \textbf{\bibinfo{volume}{9}}, \bibinfo{pages}{986} (\bibinfo{year}{2014}).

\bibitem{Veldhorst2014}
\bibinfo{author}{Veldhorst, M.} \emph{et~al.}
\newblock \bibinfo{title}{An addressable quantum dot qubit with fault-tolerant
  control fidelity}.
\newblock \emph{\bibinfo{journal}{Nature Nanotechnology}}
  \textbf{\bibinfo{volume}{9}}, \bibinfo{pages}{981} (\bibinfo{year}{2014}).

\bibitem{Kroner2008}
\bibinfo{author}{Kroner, M.} \emph{et~al.}
\newblock \bibinfo{title}{Rabi splitting and ac-stark shift of a charged
  exciton}.
\newblock \emph{\bibinfo{journal}{Applied Physics Letters}}
  \textbf{\bibinfo{volume}{92}}, \bibinfo{pages}{--} (\bibinfo{year}{2008}).
\newblock
  \urlprefix\url{http://scitation.aip.org/content/aip/journal/apl/92/3/10.1063/1.2837193}.

\bibitem{Wu1975}
\bibinfo{author}{Wu, F.~Y.}, \bibinfo{author}{Grove, R.~E.} \&
  \bibinfo{author}{Ezekiel, S.}
\newblock \bibinfo{title}{Investigation of the spectrum of resonance
  fluorescence induced by a monochromatic field}.
\newblock \emph{\bibinfo{journal}{Phys. Rev. Lett.}}
  \textbf{\bibinfo{volume}{35}}, \bibinfo{pages}{1426--1429}
  (\bibinfo{year}{1975}).
\newblock \urlprefix\url{http://link.aps.org/doi/10.1103/PhysRevLett.35.1426}.

\bibitem{Astafiev2010}
\bibinfo{author}{Astafiev, O.} \emph{et~al.}
\newblock \bibinfo{title}{Resonance fluorescence of a single artificial atom}.
\newblock \emph{\bibinfo{journal}{Science}} \textbf{\bibinfo{volume}{327}},
  \bibinfo{pages}{840--843} (\bibinfo{year}{2010}).
\newblock
  \urlprefix\url{http://www.sciencemag.org/content/327/5967/840.abstract}.
\newblock \eprint{http://www.sciencemag.org/content/327/5967/840.full.pdf}.

\bibitem{Jeschke1999}
\bibinfo{author}{Jeschke, G.}
\newblock \bibinfo{title}{Coherent superposition of dressed spin states and
  pulse dressed electron spin resonance}.
\newblock \emph{\bibinfo{journal}{Chemical Physics Letters}}
  \textbf{\bibinfo{volume}{301}}, \bibinfo{pages}{524 -- 530}
  (\bibinfo{year}{1999}).
\newblock
  \urlprefix\url{http://www.sciencedirect.com/science/article/pii/S000926149900041X}.

\bibitem{Nakamura1999}
\bibinfo{author}{Nakamura, Y.}, \bibinfo{author}{Pashkin, Y.~A.} \&
  \bibinfo{author}{Tsai, J.}
\newblock \bibinfo{title}{Coherent control of macroscopic quantum states in a
  single-cooper-pair box}.
\newblock \emph{\bibinfo{journal}{Nature}} \textbf{\bibinfo{volume}{398}},
  \bibinfo{pages}{786--788} (\bibinfo{year}{1999}).

\bibitem{DiVincenzo2000}
\bibinfo{author}{DiVincenzo, D.~P.}, \bibinfo{author}{Bacon, D.},
  \bibinfo{author}{Kempe, J.}, \bibinfo{author}{Burkard, G.} \&
  \bibinfo{author}{Whaley, K.~B.}
\newblock \bibinfo{title}{{Universal quantum computation with the exchange
  interaction.}}
\newblock \emph{\bibinfo{journal}{Nature}} \textbf{\bibinfo{volume}{408}},
  \bibinfo{pages}{339--42} (\bibinfo{year}{2000}).
\newblock \urlprefix\url{http://www.ncbi.nlm.nih.gov/pubmed/11099036}.

\bibitem{Petta2005}
\bibinfo{author}{Petta, J.~R.} \emph{et~al.}
\newblock \bibinfo{title}{{Coherent manipulation of coupled electron spins in
  semiconductor quantum dots.}}
\newblock \emph{\bibinfo{journal}{Science}} \textbf{\bibinfo{volume}{309}},
  \bibinfo{pages}{2180--4} (\bibinfo{year}{2005}).
\newblock \urlprefix\url{http://www.ncbi.nlm.nih.gov/pubmed/16141370}.

\bibitem{Hanson2007a}
\bibinfo{author}{Hanson, R.} \& \bibinfo{author}{Burkard, G.}
\newblock \bibinfo{title}{Universal set of quantum gates for double-dot spin
  qubits with fixed interdot coupling}.
\newblock \emph{\bibinfo{journal}{Phys. Rev. Lett.}}
  \textbf{\bibinfo{volume}{98}}, \bibinfo{pages}{050502}
  (\bibinfo{year}{2007}).
\newblock
  \urlprefix\url{http://link.aps.org/doi/10.1103/PhysRevLett.98.050502}.

\bibitem{Ithier2005}
\bibinfo{author}{Ithier, G.} \emph{et~al.}
\newblock \bibinfo{title}{Decoherence in a superconducting quantum bit
  circuit}.
\newblock \emph{\bibinfo{journal}{Phys. Rev. B}} \textbf{\bibinfo{volume}{72}},
  \bibinfo{pages}{134519} (\bibinfo{year}{2005}).
\newblock \urlprefix\url{http://link.aps.org/doi/10.1103/PhysRevB.72.134519}.

\bibitem{Yan2013}
\bibinfo{author}{Yan, F.} \emph{et~al.}
\newblock \bibinfo{title}{Rotating-frame relaxation as a noise spectrum
  analyser of a superconducting qubit undergoing driven evolution}.
\newblock \emph{\bibinfo{journal}{Nature Communications}}
  \textbf{\bibinfo{volume}{4}} (\bibinfo{year}{2013}).

\bibitem{Loretz2013}
\bibinfo{author}{Loretz, M.}, \bibinfo{author}{Rosskopf, T.} \&
  \bibinfo{author}{Degen, C.}
\newblock \bibinfo{title}{Radio-frequency magnetometry using a single electron
  spin}.
\newblock \emph{\bibinfo{journal}{Physical Review Letters}}
  \textbf{\bibinfo{volume}{110}}, \bibinfo{pages}{017602}
  (\bibinfo{year}{2013}).

\bibitem{Hahn1950}
\bibinfo{author}{Hahn, E.~L.}
\newblock \bibinfo{title}{Spin echoes}.
\newblock \emph{\bibinfo{journal}{Phys. Rev.}} \textbf{\bibinfo{volume}{80}},
  \bibinfo{pages}{580--594} (\bibinfo{year}{1950}).
\newblock \urlprefix\url{http://link.aps.org/doi/10.1103/PhysRev.80.580}.

\bibitem{Meiboom1958}
\bibinfo{author}{Meiboom, S.} \& \bibinfo{author}{Gill, D.}
\newblock \bibinfo{title}{Modified spin‐echo method for measuring nuclear
  relaxation times}.
\newblock \emph{\bibinfo{journal}{Review of Scientific Instruments}}
  \textbf{\bibinfo{volume}{29}}, \bibinfo{pages}{688--691}
  (\bibinfo{year}{1958}).
\newblock
  \urlprefix\url{http://scitation.aip.org/content/aip/journal/rsi/29/8/10.1063/1.1716296}.

\bibitem{Xu2012}
\bibinfo{author}{Xu, X.} \emph{et~al.}
\newblock \bibinfo{title}{Coherence-protected quantum gate by continuous
  dynamical decoupling in diamond}.
\newblock \emph{\bibinfo{journal}{Phys. Rev. Lett.}}
  \textbf{\bibinfo{volume}{109}}, \bibinfo{pages}{070502}
  (\bibinfo{year}{2012}).
\newblock
  \urlprefix\url{http://link.aps.org/doi/10.1103/PhysRevLett.109.070502}.

\bibitem{Golter2014}
\bibinfo{author}{Golter, D.~A.}, \bibinfo{author}{Baldwin, T.~K.} \&
  \bibinfo{author}{Wang, H.}
\newblock \bibinfo{title}{Protecting a solid-state spin from decoherence using
  dressed spin states}.
\newblock \emph{\bibinfo{journal}{Phys. Rev. Lett.}}
  \textbf{\bibinfo{volume}{113}}, \bibinfo{pages}{237601}
  (\bibinfo{year}{2014}).
\newblock
  \urlprefix\url{http://link.aps.org/doi/10.1103/PhysRevLett.113.237601}.

\bibitem{Kane1998}
\bibinfo{author}{Kane, B.~E.}
\newblock \bibinfo{title}{{A silicon-based nuclear spin quantum computer}}.
\newblock \emph{\bibinfo{journal}{Nature (London)}}
  \textbf{\bibinfo{volume}{393}}, \bibinfo{pages}{133--137}
  (\bibinfo{year}{1998}).
\newblock \urlprefix\url{http://fy.chalmers.se/~delsing/QI/Kane-Nature-98.pdf}.

\bibitem{Franke2015}
\bibinfo{author}{Franke, D.~P.} \emph{et~al.}
\newblock \bibinfo{title}{Mechanical tuning of ionized donor qubits in
  silicon}.
\newblock \emph{\bibinfo{journal}{arXiv preprint arXiv:1503.00133}}
  (\bibinfo{year}{2015}).

\bibitem{Barfuss2015}
\bibinfo{author}{Barfuss, A.}, \bibinfo{author}{Teissier, J.},
  \bibinfo{author}{Neu, E.}, \bibinfo{author}{Nunnenkamp, A.} \&
  \bibinfo{author}{Maletinsky, P.}
\newblock \bibinfo{title}{Strong mechanical driving of a single electron spin}.
\newblock \emph{\bibinfo{journal}{Nature Physics}}
  \textbf{\bibinfo{volume}{11}}, \bibinfo{pages}{820} (\bibinfo{year}{2015}).

\bibitem{Soykal2011}
\bibinfo{author}{Soykal, O.~O.}, \bibinfo{author}{Ruskov, R.} \&
  \bibinfo{author}{Tahan, C.}
\newblock \bibinfo{title}{Sound-based analogue of cavity quantum
  electrodynamics in silicon}.
\newblock \emph{\bibinfo{journal}{Phys. Rev. Lett.}}
  \textbf{\bibinfo{volume}{107}}, \bibinfo{pages}{235502}
  (\bibinfo{year}{2011}).
\newblock
  \urlprefix\url{http://link.aps.org/doi/10.1103/PhysRevLett.107.235502}.

\bibitem{Gustafsson2014}
\bibinfo{author}{Gustafsson, M.~V.} \emph{et~al.}
\newblock \bibinfo{title}{Propagating phonons coupled to an artificial atom}.
\newblock \emph{\bibinfo{journal}{Science}} \textbf{\bibinfo{volume}{346}},
  \bibinfo{pages}{207--211} (\bibinfo{year}{2014}).
\newblock
  \urlprefix\url{http://www.sciencemag.org/content/346/6206/207.abstract}.
\newblock \eprint{http://www.sciencemag.org/content/346/6206/207.full.pdf}.

\bibitem{Mkhitaryan2015}
\bibinfo{author}{Mkhitaryan, V.}, \bibinfo{author}{Jelezko, F.} \&
  \bibinfo{author}{Dobrovitski, V.}
\newblock \bibinfo{title}{Highly selective detection of individual nuclear
  spins using the rotary echo on an electron spin as a probe}.
\newblock \emph{\bibinfo{journal}{arXiv preprint arXiv:1503.06811}}
  (\bibinfo{year}{2015}).

\end{thebibliography}


\begin{thebibliography}{1}
\expandafter\ifx\csname url\endcsname\relax
  \def\url#1{\texttt{#1}}\fi
\expandafter\ifx\csname urlprefix\endcsname\relax\def\urlprefix{URL }\fi
\providecommand{\bibinfo}[2]{#2}
\providecommand{\eprint}[2][]{\url{#2}}

\bibitem{Itoh2014}
\bibinfo{author}{Itoh, K.~M.} \& \bibinfo{author}{Watanabe, H.}
\newblock \bibinfo{title}{Isotope engineering of silicon and diamond for
  quantum computing and sensing applications}.
\newblock \emph{\bibinfo{journal}{MRS Communications}}
  \textbf{\bibinfo{volume}{4}}, \bibinfo{pages}{143--157}
  (\bibinfo{year}{2014}).
\newblock
  \urlprefix\url{http://journals.cambridge.org/article_S2159685914000329}.

\bibitem{Pla2012}
\bibinfo{author}{Pla, J.~J.} \emph{et~al.}
\newblock \bibinfo{title}{A single-atom electron spin qubit in silicon}.
\newblock \emph{\bibinfo{journal}{Nature (London)}}
  \textbf{\bibinfo{volume}{489}}, \bibinfo{pages}{541--545}
  (\bibinfo{year}{2012}).

\bibitem{Laucht2015}
\bibinfo{author}{Laucht, A.} \emph{et~al.}
\newblock \bibinfo{title}{Electrically controlling single-spin qubits in a
  continuous microwave field}.
\newblock \emph{\bibinfo{journal}{Science Advances}}
  \textbf{\bibinfo{volume}{1}}, \bibinfo{pages}{1500022}
  (\bibinfo{year}{2015}).

\bibitem{Ithier2005}
\bibinfo{author}{Ithier, G.} \emph{et~al.}
\newblock \bibinfo{title}{Decoherence in a superconducting quantum bit
  circuit}.
\newblock \emph{\bibinfo{journal}{Phys. Rev. B}} \textbf{\bibinfo{volume}{72}},
  \bibinfo{pages}{134519} (\bibinfo{year}{2005}).
\newblock \urlprefix\url{http://link.aps.org/doi/10.1103/PhysRevB.72.134519}.

\bibitem{Yan2013}
\bibinfo{author}{Yan, F.} \emph{et~al.}
\newblock \bibinfo{title}{Rotating-frame relaxation as a noise spectrum
  analyser of a superconducting qubit undergoing driven evolution}.
\newblock \emph{\bibinfo{journal}{Nature Communications}}
  \textbf{\bibinfo{volume}{4}} (\bibinfo{year}{2013}).

\bibitem{Loretz2013}
\bibinfo{author}{Loretz, M.}, \bibinfo{author}{Rosskopf, T.} \&
  \bibinfo{author}{Degen, C.}
\newblock \bibinfo{title}{Radio-frequency magnetometry using a single electron
  spin}.
\newblock \emph{\bibinfo{journal}{Physical Review Letters}}
  \textbf{\bibinfo{volume}{110}}, \bibinfo{pages}{017602}
  (\bibinfo{year}{2013}).

\bibitem{Mkhitaryan2015}
\bibinfo{author}{Mkhitaryan, V.}, \bibinfo{author}{Jelezko, F.} \&
  \bibinfo{author}{Dobrovitski, V.}
\newblock \bibinfo{title}{Highly selective detection of individual nuclear
  spins using the rotary echo on an electron spin as a probe}.
\newblock \emph{\bibinfo{journal}{arXiv preprint arXiv:1503.06811}}
  (\bibinfo{year}{2015}).

\bibitem{Muhonen2014}
\bibinfo{author}{Muhonen, J.~T.} \emph{et~al.}
\newblock \bibinfo{title}{Storing quantum information for 30 seconds in a
  nanoelectronic device}.
\newblock \emph{\bibinfo{journal}{Nature Nanotechnology}}
  \textbf{\bibinfo{volume}{9}}, \bibinfo{pages}{986} (\bibinfo{year}{2014}).

\end{thebibliography}


\end{multicols}
\end{document}